\documentclass[12pt,a4paper,twoside]{article}
\usepackage{epsfig}
\usepackage{baltlat1}
\usepackage{wrapfig}
\begin{document}
\ \
\vspace{0.5mm}

\setcounter{page}{1}
\vspace{8mm}

\titlehead{Baltic Astronomy, vol.~12, XXX--XXX, 2003.}

\titleb{Does the interstellar magnetic field follow the Chandrasekhar-Fermi law?}

\begin{authorl}
\authorb{M. A. de Avillez}{1} and
\authorb{D. Breitschwerdt}{2}
\end{authorl}

\begin{addressl}
\addressb{1}{Department of Mathematics, University of \'Evora, R. Rom\~ao Ramalho 59, 7000 \'Evora, Portugal}

\addressb{2}{Max-Planck-Institut f\"ur Extraterrestrische Physik,
        Giessenbachstra{\ss}e, Postfach 1312, 85741 Garching, Germany}
\end{addressl}

\submitb{Received October 15, 2003}

\begin{abstract}
  We carried out 1.25 pc resolution MHD simulations of the ISM --
  including the large scale galactic fountain -- by fully tracking the
  time-dependent evolution of the magnetic field and the formation of
  shock compressed regions in a supernova-driven ISM. The simulations
  show that large scale gas streams emerge, driven by SN explosions,
  which are responsible for the formation and destruction of shocked
  compressed layers with lifetimes up to some 15 Myr. The T$\leq
  10^{3}$ K gas is distributed in filaments which tend to show a
  preferred orientation due to the anisotropy of the flow induced by
  the galactic magnetic field. The simulations also show that the
  magnetic field has a high variability, it is largely uncorrelated
  with the density and it is driven by inertial motions.  The latter
  is consistent with the fact that ram pressure dominates the flow for
  $10^{2}<$T$\leq 10^{6}$ K. For T$> 10^{6}$ K thermal pressure
  dominates, while for T$\leq 10^{2}$ K (stable branch) magnetic
  pressure takes over.

\end{abstract}


\resthead{The Interstellar magnetic field and the CF law}{M.~A.~de~Avillez \& D.~Breitschwerdt}

\sectionb{1}{INTRODUCTION}
\label{intro}
Recent work by Kim et al. (2001) and Passot \& V\'azquez-Semadeni
(2003) show somewhat contradicting results on the scaling of the
magnetic field with density in the ISM. Kim et al. using a
(200~ pc)$^{3}$ box centered in the Galactic midplane with periodic
boundary conditions, driven by SNe at a rate of 12 times the Galactic
value and using an uniform field strength of 5.8 $\mu$G orientated
along the $x-$direction, claim that the magnetic field scales as
$\rho^{0.4}$ at densities $n>1$ cm$^{-3}$, although their Fig.~2 shows
an almost order of magnitude variation in the field for the same
density. On the other hand Passot \& V\'azquez-Semadeni show
that for small Alfv\'enic Mach number the magnetic field is
uncorrelated with density, exhibiting a large scatter, which decreases
towards higher densities.

In this paper we investigate the variability of the magnetic field in
the Galactic disk and its correlation with the density in global MHD
simulations of the ISM, driven by SNe occuring at the Galactic rate.
Other important issues like the volume filling factors of the ISM
''phases'', the dynamics of the galactic fountain, the conditions for
dynamical equilibrium and the importance of convergence of these
results with increasing grid resolution have been treated elsewhere
(Avillez 2000, Avillez \& Breitschwerdt 2004a, b).

\sectionb{2}{MODEL AND SIMULATIONS}

We ran kpc-scale MHD simulations of the SN-driven ISM on a cartesian
grid of $0\leq (x,y)\leq 1$ kpc size in the Galactic plane and $-10
\leq z \leq 10$ kpc into the halo with a finest adaptive mesh
refinement resolution of 1.25 pc, using a modified version of the 3D
model of Avillez (2000). The new model allows for the motion of OB
associations and includes magnetic fields (Avillez \& Breitschwerdt
2004b). At the beginning of the simulations the uniform field
components along the three axes are given by $(B_{u,0}
(n(z)/n_0)^{1/2},0,0)$, where $B_{u,0}=3~\mu$G is the field strength,
$n(z)$ is the number density of the gas as a function of distance from
the Galactic midplane and $n_0=1 \, {\rm cm}^{-3}$ is the average
midplane density.  The random field component is set to zero in the
beginning of the simulations. This component is built up during the
first millions of years of evolution as a result of turbulent motions,
mainly induced by SN explosions.

\begin{figure}[t]
\vspace*{1.in}

\hspace*{0.1\hsize}Left panel: mavillez$\_$fig1a, Right panel: mavillez$\_$fig1b
\vspace*{1.in}

\captionb{1}{Density and magnetic field distribution in the
Galactic midplane after 374 Myr of evolution. The resolution of
the finest AMR level is 1.25 pc.} \label{fig1}
\end{figure}

\sectionb{3}{RESULTS}
\subsectionb{3.1}{Global Evolution}
The initial evolution of the magnetized disk is similar to that seen
in purely HD runs (Avillez 2000, Avillez \& Breitschwerdt 2004a), that
is, the initially stratified distribution does not hold for long as a
result of the lack of equilibrium between gravity and (thermal,
kinetic and turbulent) pressure during the ``switch-on phase'' of SN
activity. As a consequence the gas in the upper and lower parts of the
grid collapses onto the midplane, leaving low density material in its
place. However, in the MHD run it takes a longer time for the collapse
to be completed as a result of the opposing magnetic pressure and
tension forces. As soon as the system has collapsed and enough
supernovae have gone off in the disk building up the required pressure
support, transport into the halo is not prevented, although the escape
of the gas takes a few tens of Myr to occur.  The crucial point is
that a huge thermal overpressure due to combined SN explosions can
sweep the magnetic field into dense filaments and punch holes into the
extended warm and ionized H{\sc i} layers. Once such pressure release
valves have been set up, there is no way from keeping the hot
over-pressured plasma to follow the pressure gradient into the halo.
As a consequence the duty disk-halo-disk cycle of the hot gas is fully
established, in which the competition of energy input and losses into
the ISM by SNe, diffuse heating, radiative cooling and magnetic
pressure leads the system to evolve into a dynamical equilibrium state
within a few hundred Myr.  This time scale is considerably longer than
that quoted in other papers (e.g., Korpi et al. 1999, Kim et al.
2001), because in these the galactic fountain has not been taken into
account.

Fig.~1 shows slices of the 3D data cube of the density
and magnetic field distributions in the Galactic midplane. The highest
density gas tends to be confined to shocked compressed layers that
form in regions where several large scale streams of convergent flow
(driven by SNe) occur. The compressed regions, which have on average
lifetimes of 10-15 Myr, are filamentary in structure, tend to be
aligned with the local field and are associated with the highest field
strengths. The formation time of these high density structures depends
on how much mass is carried by the convergent flows, how strong the
compression is and on the rate of cooling of the regions under
pressure.

\subsectionb{3.2}{Field Variability and Dependence with Density}
\begin{wrapfigure}{i}[0pt]{64mm}
\vspace*{0.95in}
\hspace*{0.25\hsize}mavillez$\_$fig2.jpg
\centerline{}
\vspace*{1in}

  \captionb{2}{Scatter plot of B vs $\rho$ for the T$\leq 10^{3}$
    (black), $10^{3}<$T$\leq 10^{4}$ (green), $10^{4}<$T$\leq10^{5.5}$
    K (blue) and T$>10^{5.5}$ K (red) regimes at 400 Myr of disk
    evolution.}
\end{wrapfigure}

During the evolution of the system, thermal and dynamical processes
broaden the distribution of the field strength in such a way that
after the global dynamical equilibrium has been set up the field
strength in the disk spans two orders of magnitude from $10^{-7}$ to
$10^{-5}$ G (Fig.~2). The figure shows a \emph{large}
scatter in the field for the same density suggesting that the field is
being driven by the inertial motions, rather than it being the agent
determining the motions. In the latter case the field would not be
strongly distorted, and it would direct the motions predominantly
along the field lines. The high field variability is also seen in the
right panel of Fig.~1, which shows a highly turbulent field, that
seems to be uncorrelated with the density, and thus, the classical
scaling law $B\sim\rho^\alpha$, with $\alpha=1/2$, according to the
Chandrasekhar-Fermi (CF) model (1953) will not hold. It should be kept
in mind that in CF it was assumed that the field is distorted by
turbulent motions that were subalfv\'enic, whereas in our simulations
in addition both supersonic and superalfv\'enic motions can occur,
leading to strong MHD shocks. It now depends if the density
fluctuations are primarily caused by thermal and/or ram pressure
fluctuations and therefore being uncorrelated with the magnetic field,
or if MHD waves are the driving agent and provide a coupling between
matter and field. Therefore, in general $0\leq \alpha \leq 1$ would be
expected. However, it should be noted that in reality heating and
cooling processes, and even magnetic reconnection could induce further
changes. Observationally there seems to be some evidence from
measuring magnetic field strengths in the cold neutral medium, that
$B$ and $\rho$ can be largely uncorrelated (Troland \& Heiles 2001).

\begin{figure}[t]
  \includegraphics[width=0.5\hsize,angle=0]{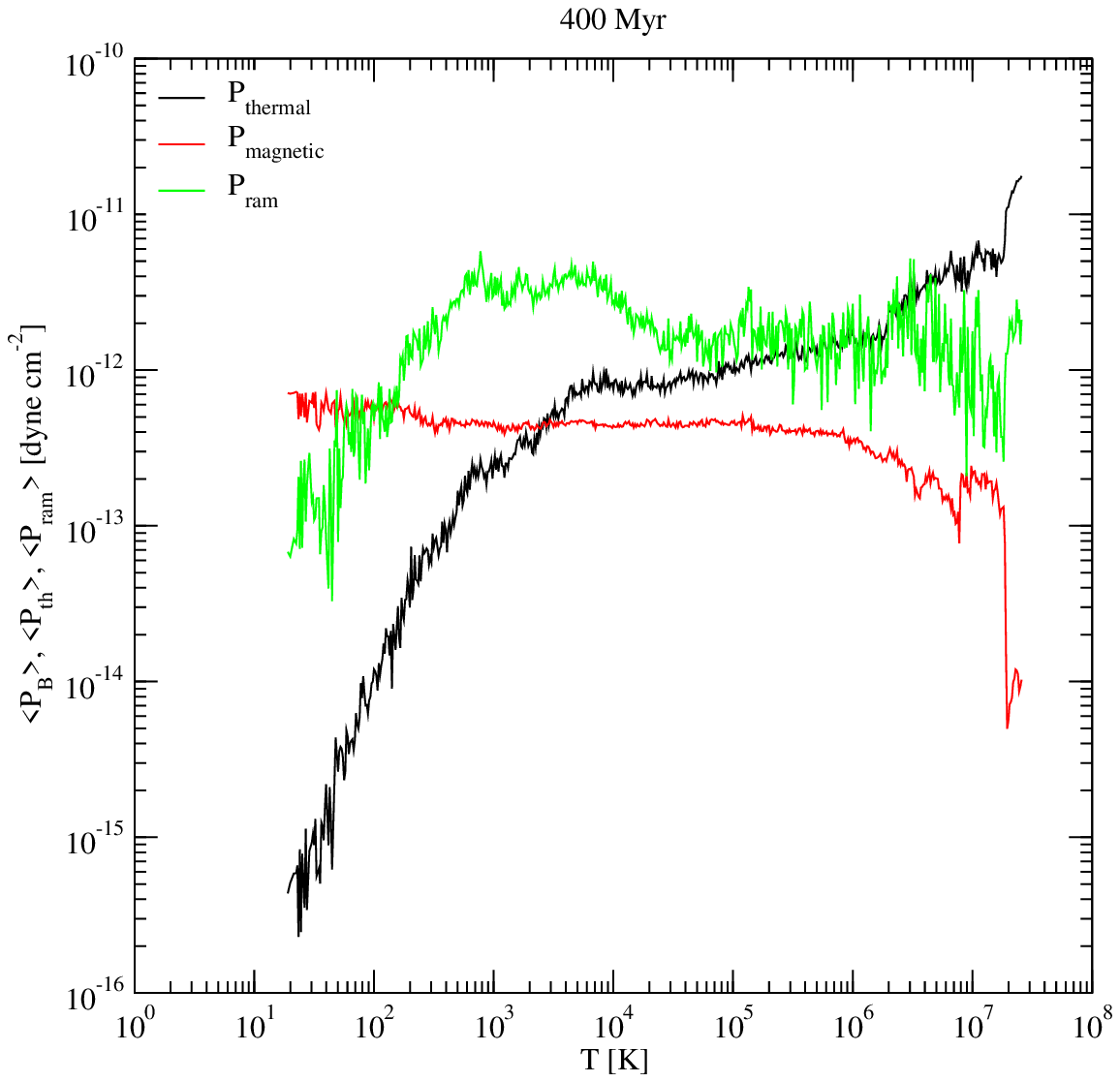}
  \includegraphics[width=0.5\hsize,angle=0]{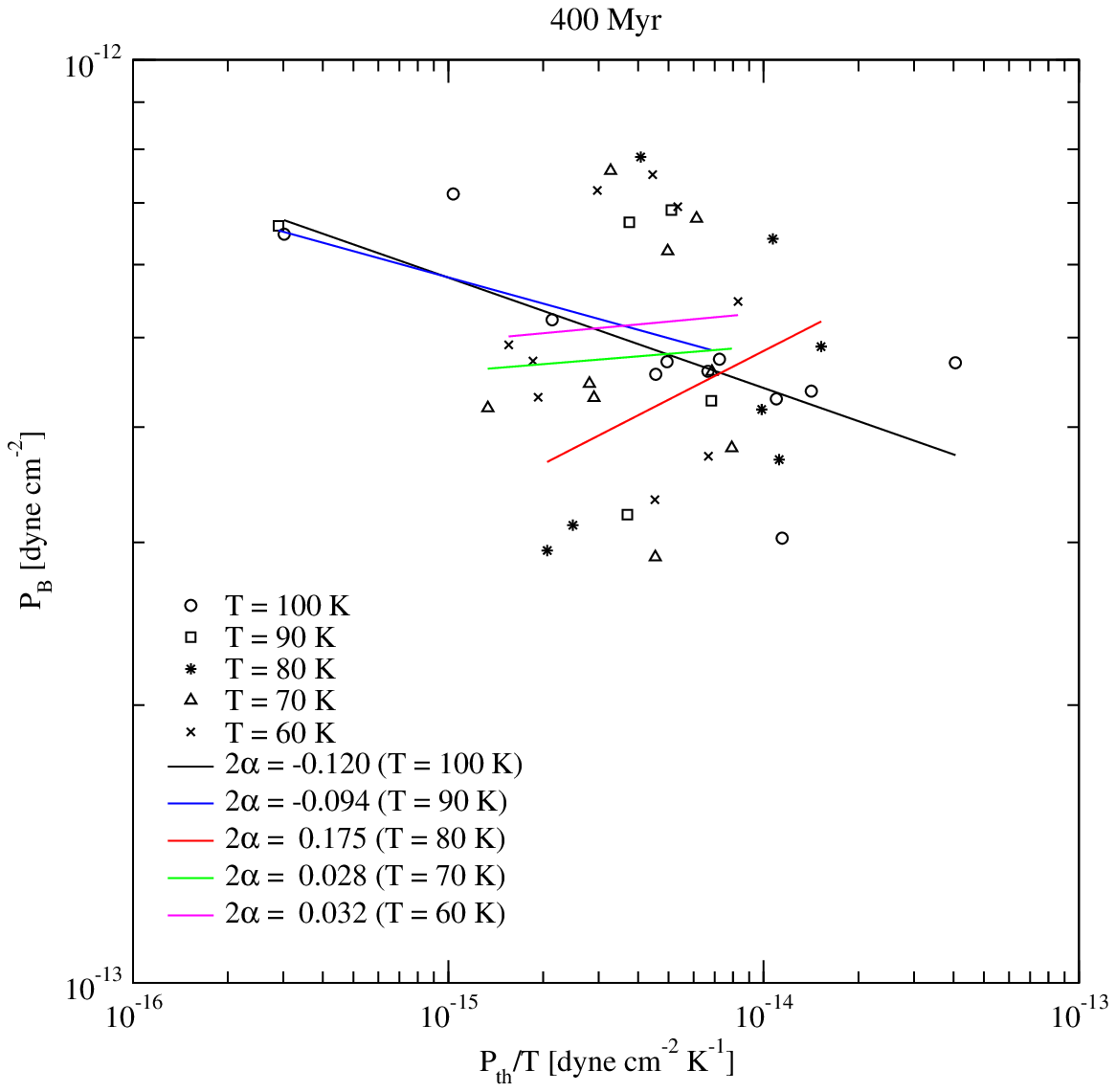}
  \captionb{3}{The left panel shows the average magnetic (red),
    thermal (black) and ram (green) pressures as functions of
    temperature at 400 Myr. The right panel shows the variation of
    P$_{B}$ with P$_{th}/$T for T$<10^{2}$ K gas in the disk.  The
    $2\alpha$ parameters are the slopes of the straight lines and
    correspond to the relation P$_{B}\propto
    \left(\mbox{P}_{th}/\mbox{T}\right)^{2\alpha}$ if a $B\propto
    \rho^{\alpha}$ law is assumed.}
\end{figure}

The left panel of Fig.~3 shows that the T$\leq 10^{2}$ K gas has
$\mbox{P}_{B}>\mbox{P}_{ram}\gg\mbox{P}_{th}$, demonstrating that
magnetically dominated regions do exist, while the T$> 10^6$ K gas has
the highest thermal pressure and the lowest magnetic pressure. For
T$<10^{2}$ K the relation $B \propto \rho^{\alpha}$ does not hold
either, as $\alpha$ varies between -0.006 and 0.085 (right panel of
Fig.~3) suggesting that the thermal and magnetic pressures are
independent. This result is largely consistent with the almost zero
variation of magnetic pressure with temperature seen in the left panel
of Fig.~3. For $10^{2}<$T$<10^{6}$ K ram pressure determines the
dynamics of the flow, and therefore, the magnetic pressure does not
act as a restoring force (Passot \& V\'azquez-Semadeni 2003) as it was
already suggested by the lack of correlation between the field
strength and the density. Fig.~3 also shows that the basic assumption
of energy equipartition, made in the paper by CF (1953) in order to
calculate the magnetic field in the spiral arm, is clearly not
fulfilled.

\sectionb{4}{DISCUSSION}

The dynamical picture that emerges from these simulations is that
thermal pressure gradients dominate mostly in the neighborhood of SNe,
which drive motions whose ram pressures are dominant over the mean
thermal pressure (away from the energy sources) and the magnetic
pressure. The magnetic field is only dynamically important at low
temperatures, but can also weaken gas compression in MHD shocks and
hence lower the energy dissipation rate. The thermal pressure of the
freshly shock heated gas exceeds the magnetic pressure by usually more
than an order of magnitude and the B-field can therefore not prevent
the flow from rising perpendicular to the galactic plane.  Thus hot
gas is fed into the galactic fountain at almost a similar rate than
without field.

The present simulations do not include self-gravity, which could
have an important effect on high density regions, although on the
verge of Jeans instability these would decouple from the ambient
ISM flow. Moreover heat conduction has been neglected in the view
of it being second order in comparison to the dominance of bulk
motions and turbulent mixing. In the spirit of building a
bottom-up model of the ISM further components (e.g., cosmic rays)
and processes (e.g., non-equilibrium cooling) will be included in
future simulations.

ACKNOWLEDGMENTS.\ The authors acknowledge fruitful discussions with E. V\'asquez-Semadeni and M.-M. Mac Low.
\goodbreak

\References
\ref 
Avillez M.~A. 2000, MNRAS, 315, 479

\ref 
Avillez M.~A., Breitschwerdt~D. 2004a, A\&A, in press

\ref 
Avillez M.~A., Breitschwerdt~D. 2004b, Ap\&SS, in press (Astro-ph/0310634)

\ref
Chandrasekhar~S., Fermi~E. 1953, ApJ, 118, 113

\ref 
Ferri\`{e}re K.~M. 1998, ApJ, 503, 700

\ref 
Kim~J., Balsara~D., Mac Low~M.-M. 2001, JKAS, 34, S333

\ref 
Korpi~M.~J., Brandenburg~A., Shukurov~A., Tuominen~I., \& Nordlund~A. 1999, ApJ, 514, L99

\ref 
Passot~T., V\'azquez-Semadeni~E. 2003, A\&A, 398, 845

\ref 
Troland~T.~H., Heiles~C., 2001, BAAS, 33, 918
\end{document}